\documentclass[10pt]{article}

\usepackage{amsfonts,amsbsy,graphicx}
\usepackage{natbib}
\usepackage[colorlinks=true,urlcolor=blue,linkcolor=blue,citecolor=blue]{hyperref}

\newcommand{\C}{\mathbb{C}}
\newcommand{\R}{\mathbb{R}}
\newcommand{\ovl}{\overline}
\newcommand{\reta}[1]{\lower-4pt\hbox{${\displaystyle\longleftrightarrow\atop{\displaystyle #1}}$}}

\begin{document}
\centerline{\LARGE{\bf An interactive programme for Steiner trees}}
\ \\
\ \\
\centerline{\Large{Marcelo Zanchetta do Nascimento$^1$, Val\'erio Ramos Batista$^2$,}}\
\centerline{\Large{Wendhel Raffa Coimbra$^3$}}
\ \\
\ \\
\centerline{\footnotesize{$^1$ CMCC-UFABC, r. Sta. Ad\'elia 166, 
Bl.B, 09210-170 St. Andr\'e-SP, Brazil}}
\centerline{\footnotesize{{\it Tel:} +55-11-4996-0077, {\it Email}: {\tt marcelo.zanchetta@gmail.com}}}
\centerline{\footnotesize{$^2$ CMCC-UFABC, r. Sta. Ad\'elia 166, 
Bl.A, 09210-170 St. Andr\'e-SP, Brazil}}
\centerline{\footnotesize{$^3$ UFMS, rod. BR 497 km 12, 79500-000 Paranaiba-MS, Brazil}}
\bigskip

{\Large \textbf{Abstract}}\\[4mm]
\fbox{
\begin{minipage}{5.4in}{\footnotesize We introduce a fully written programmed code with a supervised method for generating Steiner trees. Our choice of the programming language, and the use of well-known theorems from Geometry and Complex Analysis, allowed this method to be implemented with only 747 lines of effective source code. This eases the understanding and the handling of this beta version for future developments.} 
\end{minipage}}\\[1mm]

\section{Introduction}
\label{intro}

One of the main problems at implementing multicast in Wide Area Networks (WAN) is the high cost of transmissions between terminals. Cost reduction is attained by adding routers to the network, but this increases complexity (see \cite{J,Sa}). Steiner trees have long been used in order to optimise routes, aiming at the lowest cost possible (see \cite{A,H}). 
\\

However, the Steiner tree problem belongs to the NP-hard class (see \cite{GGJ}). For generating minimal Steiner trees, non-supervised methods must then either be restricted to a few number of points or make use of heuristics (see \cite{DZ,RS}; \cite{S}; \cite{Wa}; \cite{Wi1,Wi2,WC}; \cite{FM}; \cite{YG} for examples). See also
\\

{\tt http://www.diku.dk/hjemmesider/ansatte/martinz/geosteiner}
\\
\\
for a heuristic-based open source code.  
\\

Therefore, if one really seeks the {\it minimal} Steiner tree, there are little chances that non-supervised methods will find it. Of course, a supervised method that includes some feedback to the user increases the chances of finding this tree. Specially if we can rely on the skill and good guesses of a trained user. By the way, in the last paragraph from \S 6 of \cite{GP} the very authors had already made this kind of comment.
\\

In this same work, Gilbert and Pollak conjectured that a Steiner minimal tree (SMT) must have its length in the interval $Lprim\cdot[\sqrt{3}/2,1]$, where {\it Lprim} is the length of the minimal spanning tree. This one can be obtained by Prim's algorithm \cite{P}, and here we shall name it after him for concision and clarity. According to \cite{DH}, this conjecture is right. If so, any Steiner tree obtained from Prim's can improve it of at most 13.4\%, let us say ``6.7\% in general''. Therefore, our programme could help find trees up to 1-0.866/(1-0.067)=7.2\% better than Steiner trees constructed by many algorithms in the literature.
\\

It seems to be little, but if a connection is used extensively, this 7.2\% will represent a great economy in the long term. Moreover, in \cite{IK} the authors claim that the Gilbert and Pollak's conjecture is not completely answered yet. Thus, it might even happen that we get an SMT {\it under} this ratio. The programme can help users who are convinced of this possibility, and their trained intuition could even find a possible counterexample. 
\\

Of course, a supervised programme is not suitable for thousands of terminals, except for a long term project distributed to a team of users. But even this exception does not apply for extreme cases, like VLSI-design through {\it rectilinear} Steiner trees, in which {\it millions} of terminals are needed. However, terminals in hundreds are still the case in several applications, like sound and video cards. Their frequent access makes desirable to minimise delay as much as possible. Even a 1\% improvement would count in this case.
\\

We introduce a fully written programmed code for generating Steiner trees. Our choice of a programming language was made in order to spare the graphical environment, which Matlab already brings in a very well built-in way. Since we have not used toolboxes, this code can be easily adapted to a free software like Octave, though this one has a simpler graphical user interface. Anyway, this produces a much shorter code, easier to understand and to handle for future developments. Indeed, the present version of our programme has didactic purposes, for it is accessible even to graduate students in Computer Science.
\\

As we mentioned above, the tree of Prim is frequently used to construct a Steiner tree. But depending on the {\it way} this construction works, we get a non-minimal Steiner tree. See Figures 1 and 2 for an example. We recall that numerics must use a tolerance for theoretical angles. In our case, Steiner points are vertices of angles that measure $120^\circ\pm 2^\circ$. 
\\
\input epsf
\begin{figure}[ht]
\epsfxsize 7cm  
\epsfbox{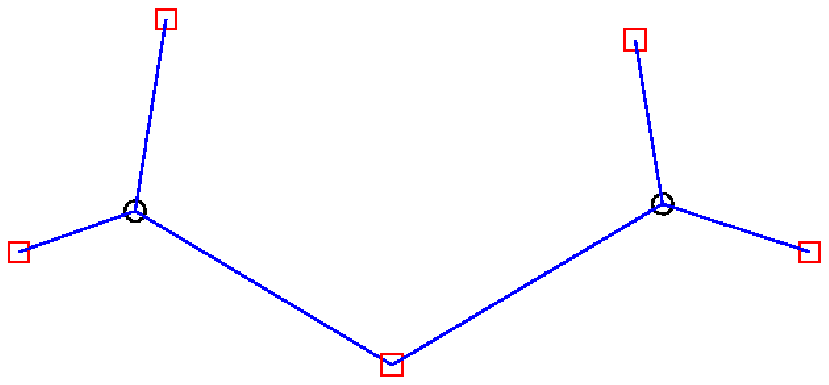}
\epsfxsize 7cm  
\epsfbox{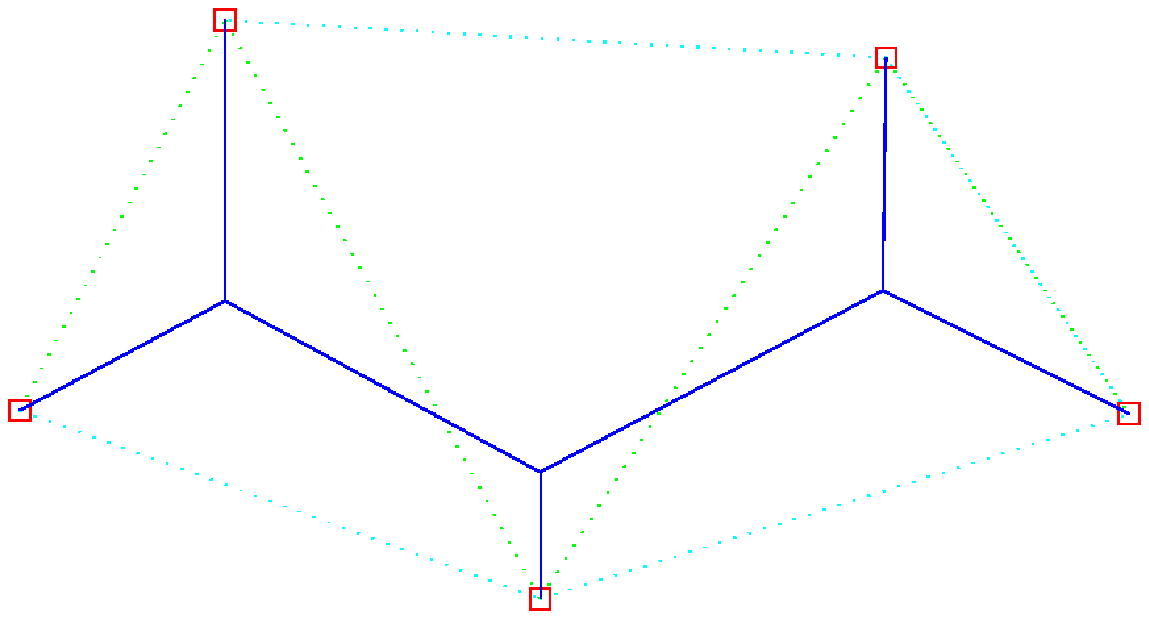}
\centerline{Figure 1: Steiner tree out of Prim's.\hfill Figure 2: Steiner minimal tree.}
\end{figure}

Many {\it original} ideas were used to write this code. They are based on Theorems normally relegated to Math courses only, specially from Geometry and Complex Analysis. But the chosen programming language allows these theoretical results to be implemented into elegant and efficient codes. The reader can download some of them from 
\\

{\tt http://hostel.ufabc.edu.br/\~{}marcelo.nascimento/software.html}
\\
\\
and the full programme in p-code at the same address in order to make tests. As an example, the file ``Prim.m'' runs with only 14 lines of source code. In total, {\tt stree.m} and its related to programmes have only 822 lines, which drop to 747 if we do not count {\tt sread.m} and {\tt swrite.m} \ (for graphical input and output of terminal points).  
\\

The aim of this paper is to introduce the programme {\tt stree.m}, totally written with original ideas we have just mentioned in the previous paragraph. The rest of this paper is organised as follows: in Section \ref{gs} we briefly describe how the programme works. It draws full tree stretches automatically with the mouse, and Section \ref{fts} gives details and hints about it. Section \ref{st} is devoted to explaining some theorems that we used to write the programme. In Section \ref{ec} we expose some of the geometrical and analytical ideas used to implement {\tt stree.m}, and make our conclusions in Section \ref{ccls}. The full source code will be available in future.

\section{Getting started}
\label{gs}

This present section is like a quick manual to the programme. The readers that are more interested in mathematical results can read it through or simply go straight to Section \ref{fts}.   
\\

Download {\tt stree.zip} from {\tt http://hostel.ufabc.edu.br/\~{}marcelo.nascimento/software.html} and extract it in a folder. Enter ``stree'' at the Matlab prompt. You will get the following message:
\\

{\sf Please adjust terminal window to show full picture.

Give a filename to open or press Enter to choose points.}
\\

Adjusting the terminal window will prevent figures to hide it. Now you may either give an existing datafile of terminal points for the tree or plot one with the mouse left-button. Some test-files with extension ``.txt'' can be found in {\tt stree.zip}, and we shall take them here for comments. 
\\

Press the Enter key after you either finish plotting points or type the filename {\it without} extension, for instance {\tt test0}. Figure 3 will appear with the tree of Prim, and Figure 4 is for you to draw a Steiner tree. Our purpose is to find the {\it shortest} tree, and Figure 3 can help make good choices.
\\
\input epsf
\begin{figure}[ht]
\epsfxsize 7cm  
\epsfbox{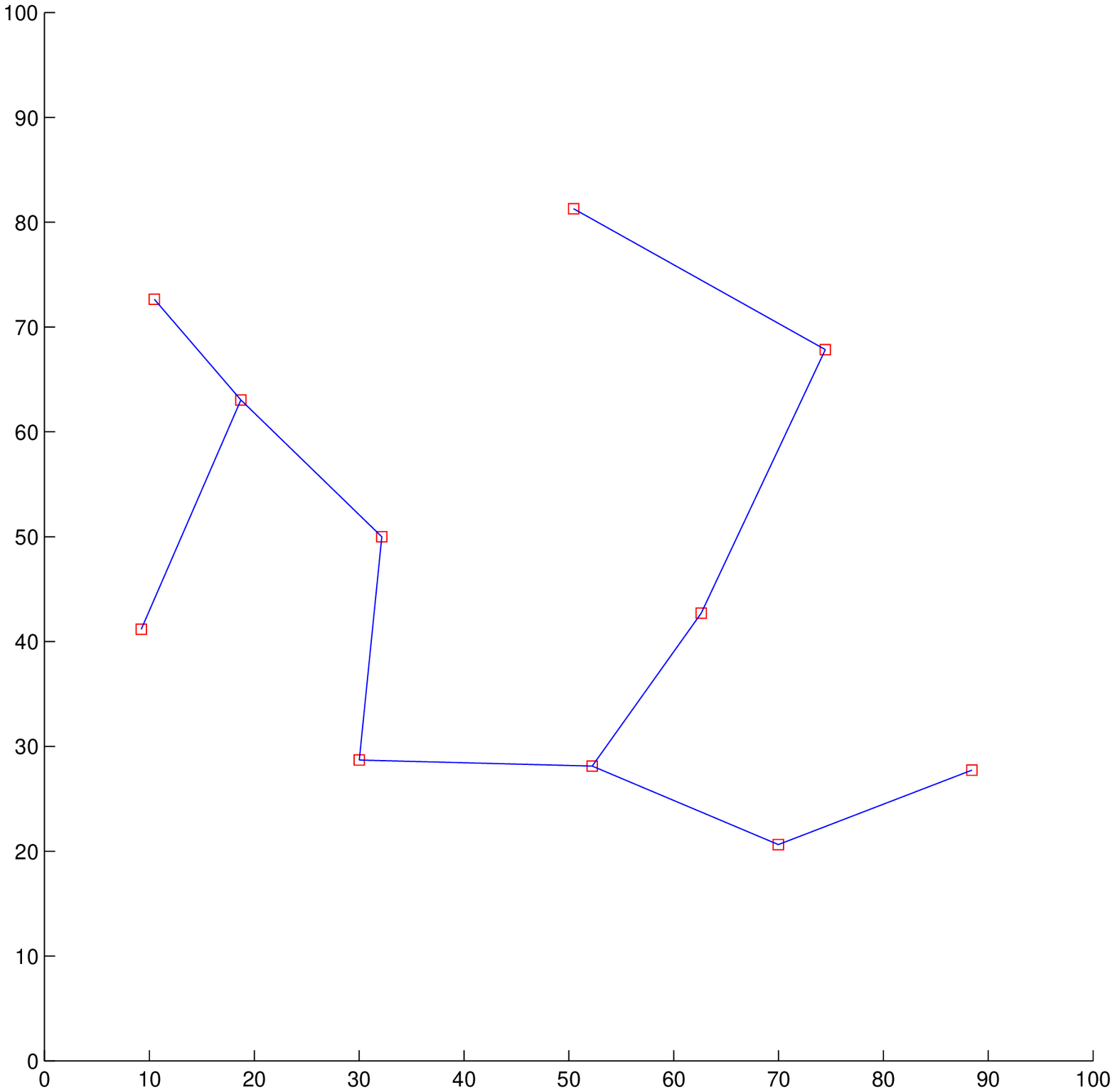}
\epsfxsize 7cm  
\epsfbox{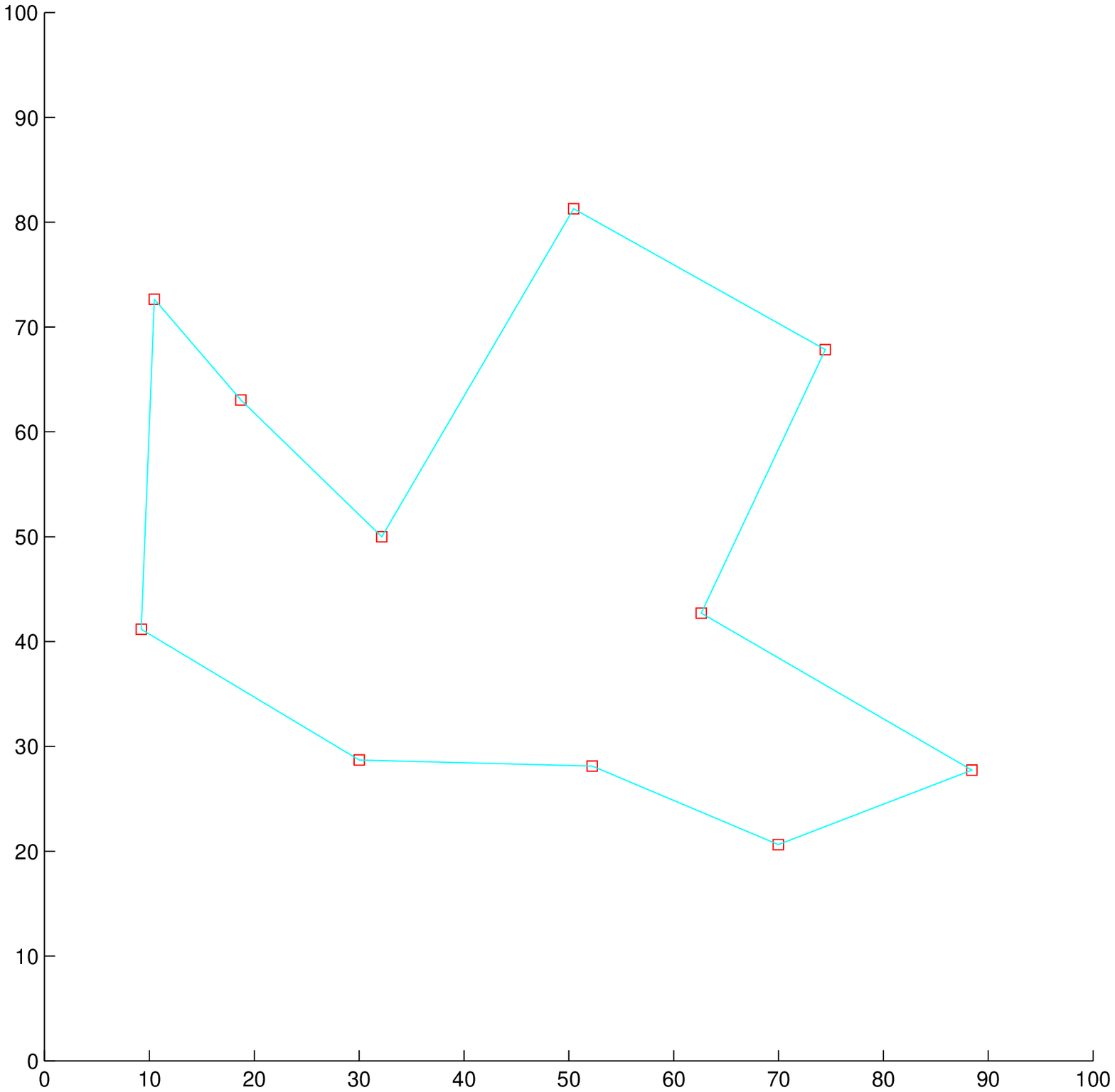}
\centerline{Figure 3: The tree of Prim.\hfill Figure 4: Terminals joined by auxiliary cyan edges.}
\end{figure}

In Figure 4, the cyan polygon is obtained by the Convex Hull and Lune properties described in \cite{GP}. It is called ``Steiner hull''. From this point on, some hints and instructions to proceed are printed in the Matlab terminal window, and all actions will be with the mouse. For the tree of Prim, its length {\it Lprim} and $\frac{\sqrt{3}}{2}Lprim$ are printed as follows:
\\

{\sf The length of a SMT ranges from/to

ans =

182.8525

Lprim =

211.1398}
\\
\\
and so the user gets an estimate on the length of an SMT from Gilbert and Pollak's conjecture. However, according to \cite{IK}, it might even happen that we get an SMT {\it under} this ratio.
\\

From the {\it first mouse menu}, now printed in the Matlab terminal window, you can read the drawing options:
\\

{\sf Press button

l(+l)+r to omit point(s);

r alone to try full tree;

middle  for more options.}
\\

For instance, choose $(75,68)$ as the first and $(30,29)$ as the second point with the left button, and then press the right button. A dotted black polygonal will blink to indicate a remaining cyan polygon, of which the vertices are now joined by a Steiner subtree computed by the programme (see Figures 5 and 6). 
\\

From the first to the second point, one goes {\it counterclockwise} along the black polygon. It is a standard of {\tt stree.m}, but if you mistake it {\it before} clicking the right button, choose extra point(s) and the programme will treat the latter two as ``first'' and ``second''. Otherwise, if you mistake that standard and realise it only {\it after} having clicked the right button, it is again no problem. The first menu is re-printed on the screen, and now you can undo your previous step as follows: press the middle button for the {\it second mouse menu}, namely:
\\

{\sf Press button

left   to return drawing;

right  to undo previous step;

middle for retouches.}
\\

From it we have that the right button now undoes your mistake, and with the left button you will be back again to the first menu.
\\
\input epsf
\begin{figure}[ht]
\epsfxsize 7cm  
\epsfbox{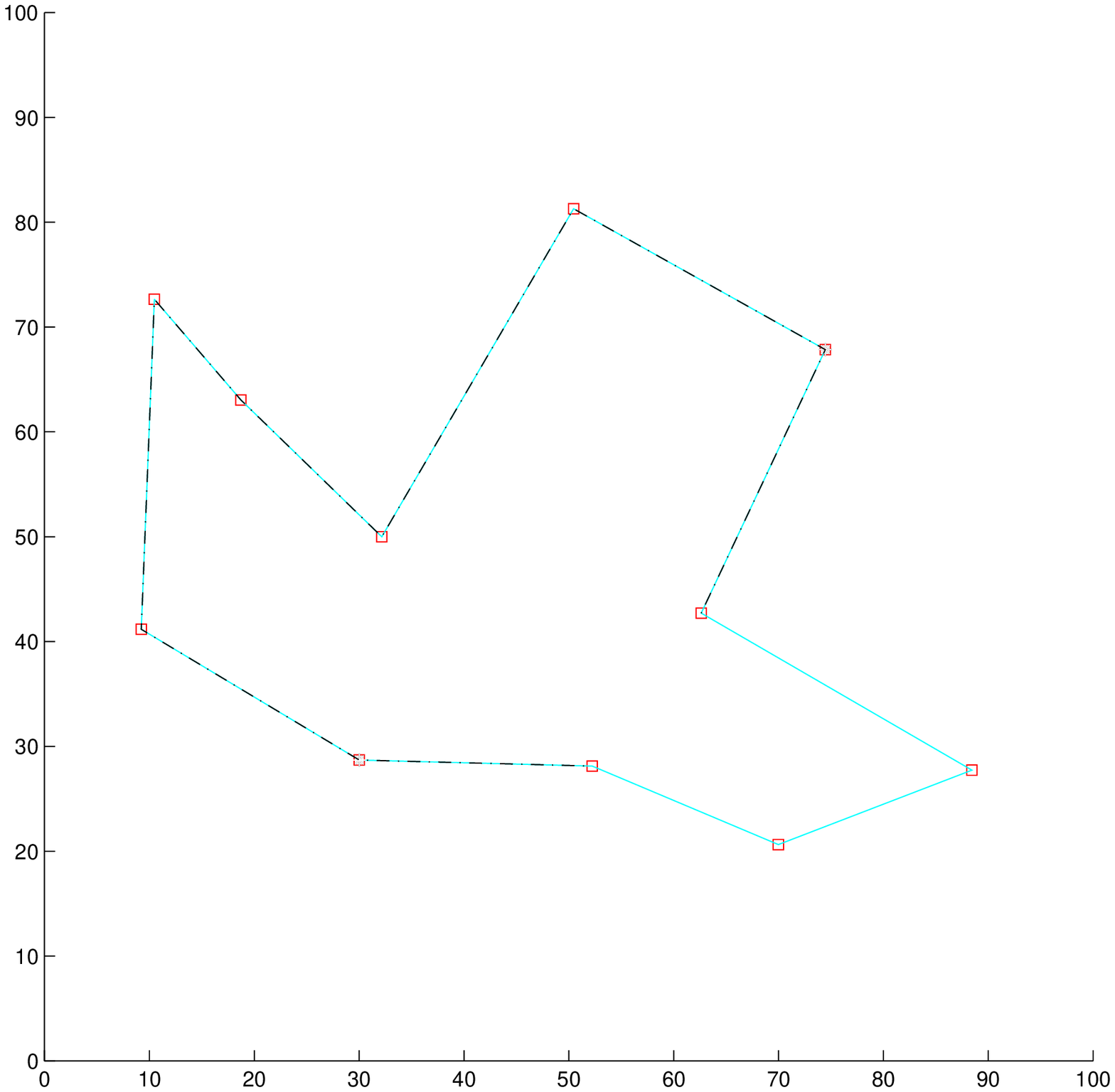}
\epsfxsize 7cm  
\epsfbox{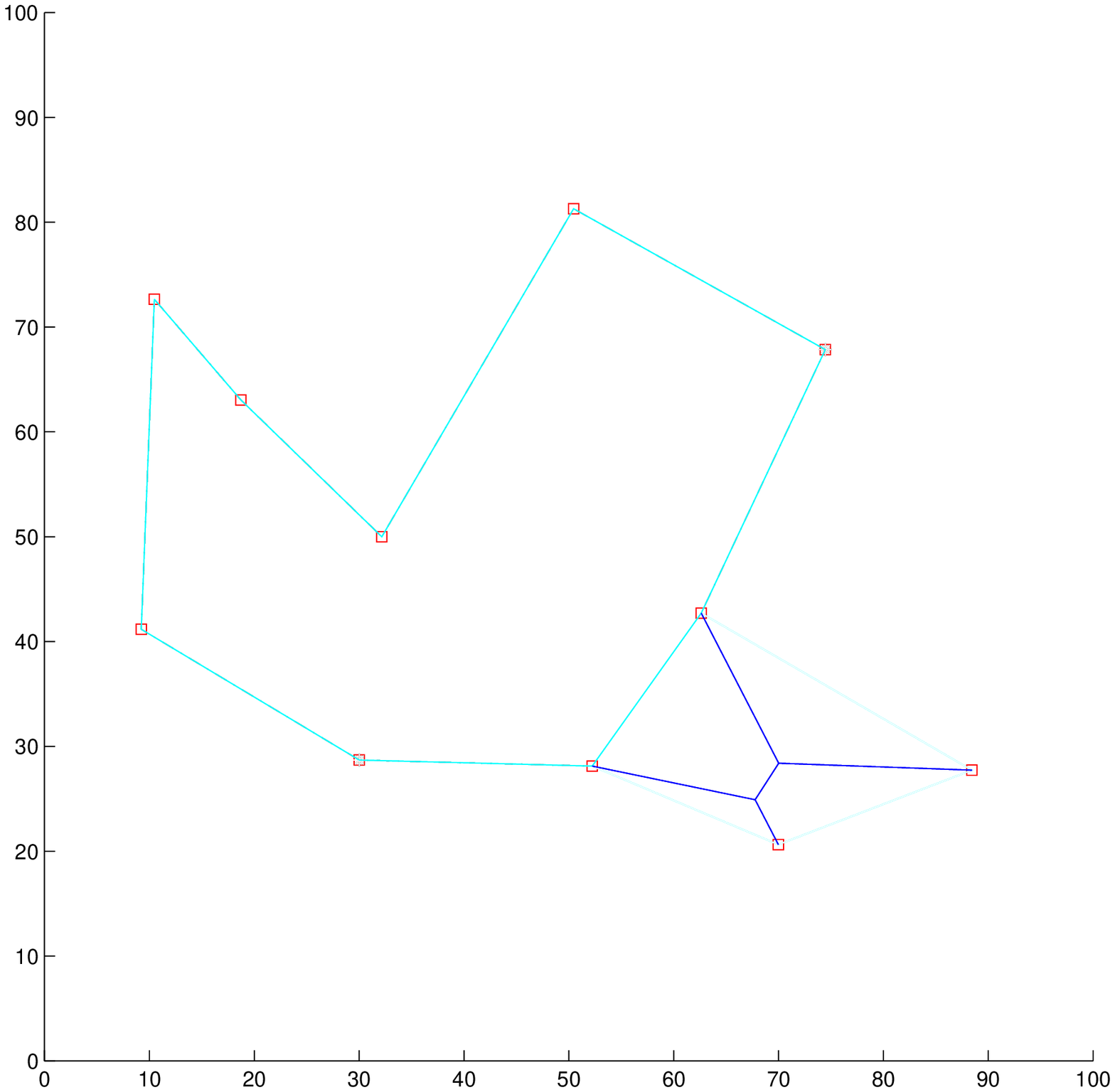}
\centerline{Figure 5: The dotted black polygonal.\hfill Figure 6: A Steiner subtree.}
\end{figure}

You might now want to take $(19,63)$ and then $(52,29)$ to get a Steiner subtree out of a quadrilateral. That will work out, but Figure 3 does not hint this way. So try instead $(32,50)$ and $(52,29)$ and you will get Figure 7 with an ``odd'' cyan polygon. However, {\tt stree.m} shows it for you to consider other possibilities not hinted by Figure 3, for instance Figure 8. But in this figure, the Steiner tree is not minimal because it has length {\it Ltree} $=227.1818$, which is bigger than {\it Lprim} $=211.1398$, as printed in the Matlab terminal window. Also, a full border is printed when you terminate execution.
\\
\input epsf
\begin{figure}[ht]
\epsfxsize 7cm  
\epsfbox{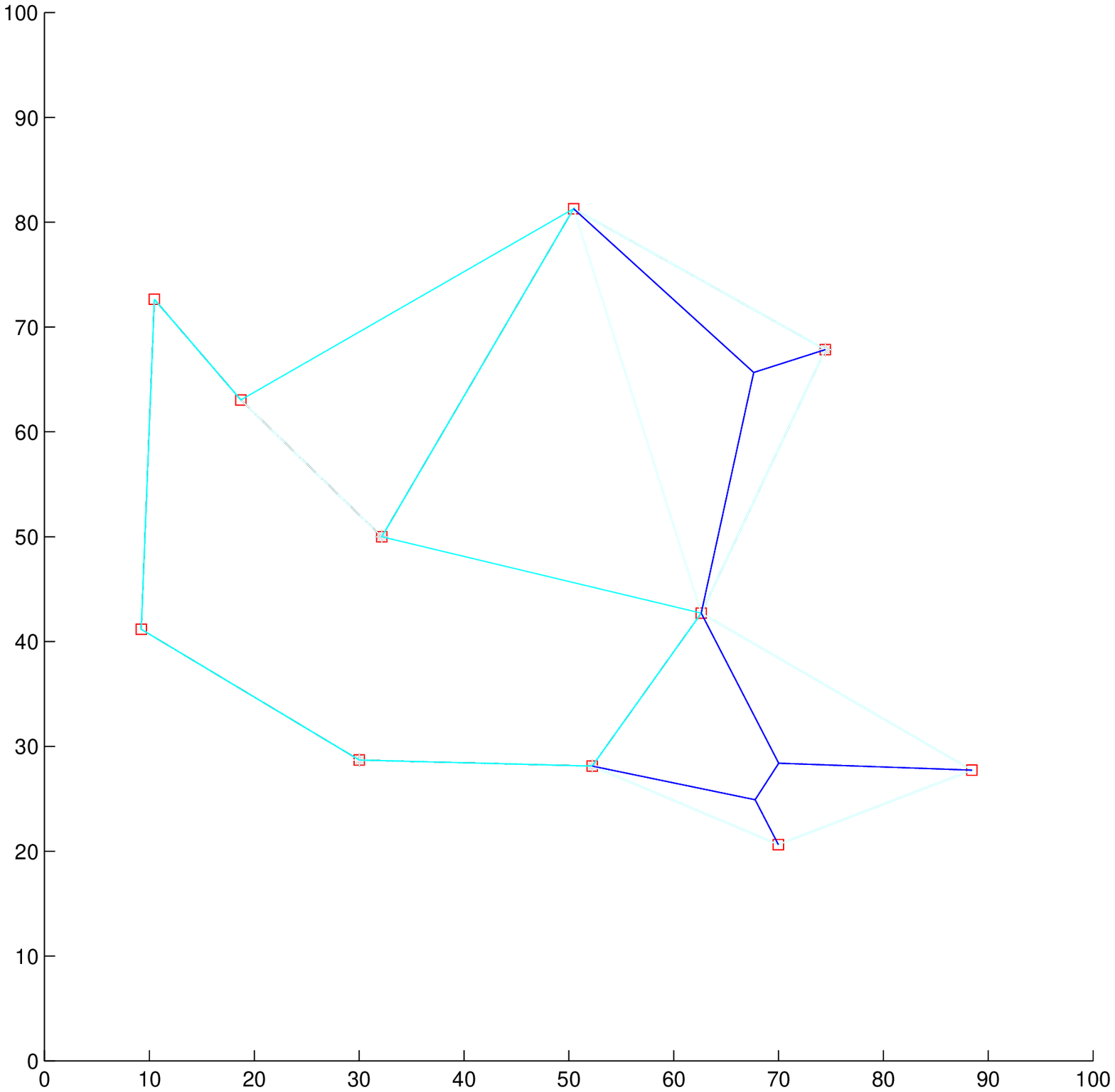}
\epsfxsize 7cm  
\epsfbox{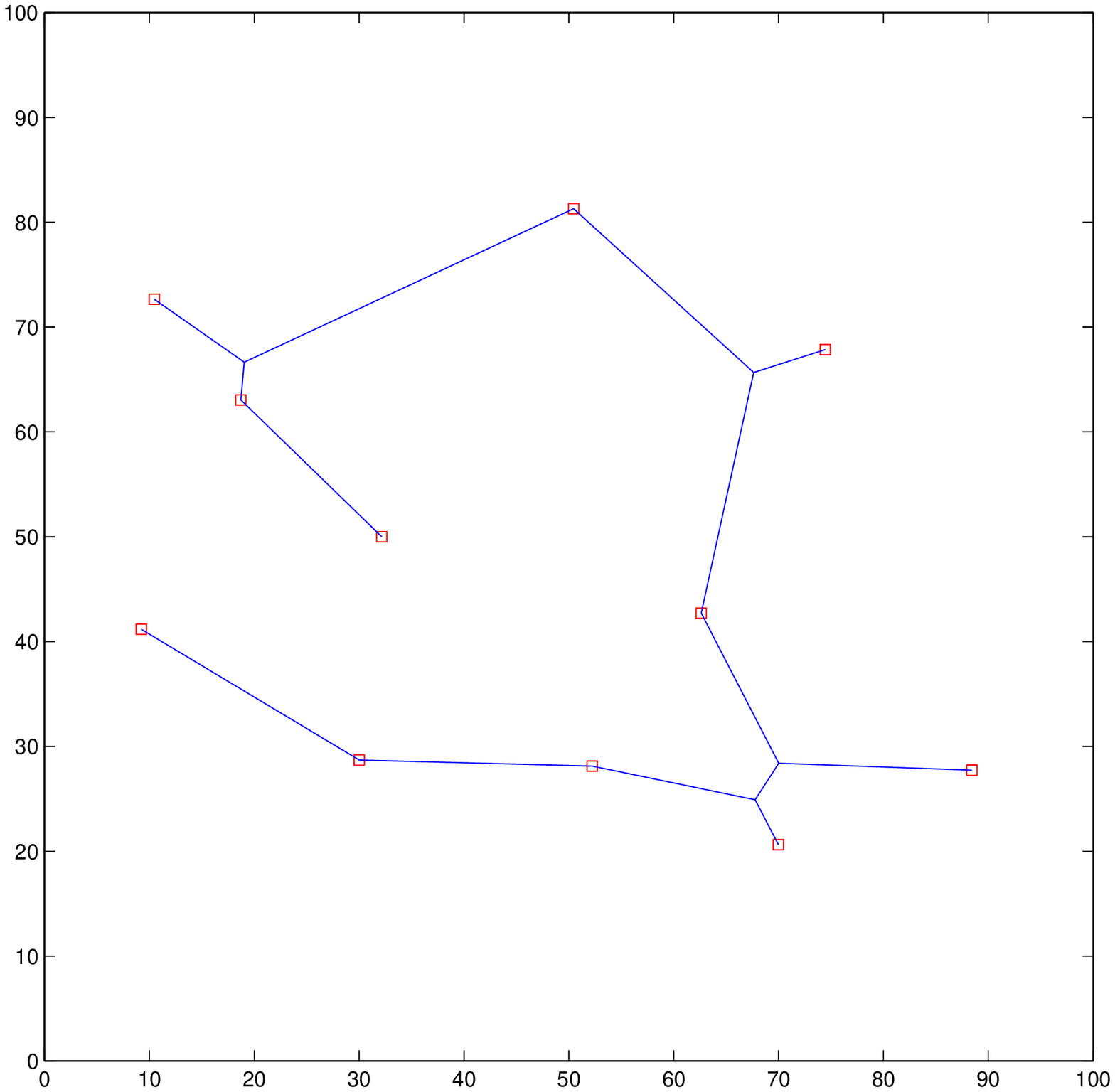}
\centerline{Figure 7: An ``odd'' cyan polygon.\hfill Figure 8: A non-minimal Steiner tree.}
\end{figure}

Back to Figures 7 and 3, remove now $(50,81)$ by clicking on it with the left and then right mouse buttons. Remove $(10,73)$, and then $(62,43)$ in the same way. The order you remove these single points is unimportant. You will then be left with a group of terminals very likely to give a full Steiner subtree. It will be drawn by pressing the right mouse button alone, no matter the cursor is. The Steiner tree will be concluded by connecting the isolated point $(10,73)$. By the {\it third mouse menu}, 
\\

{\sf Thrice Left = Steiner Point

Left + n x Right = Polygonal

Middle = Terminate}
\\
\\
which is also the last one, you should click left on $(19,63)$ and then right on $(10,73)$. The middle mouse button concludes your drawing (see Figure 9) and informs that {\it Ltree} $=206.1456$. Sometimes, as in this case, you go from the first to the third mouse menu without access to the second. 
\\
\input epsf
\begin{figure}[ht]
\epsfxsize 7cm  
\epsfbox{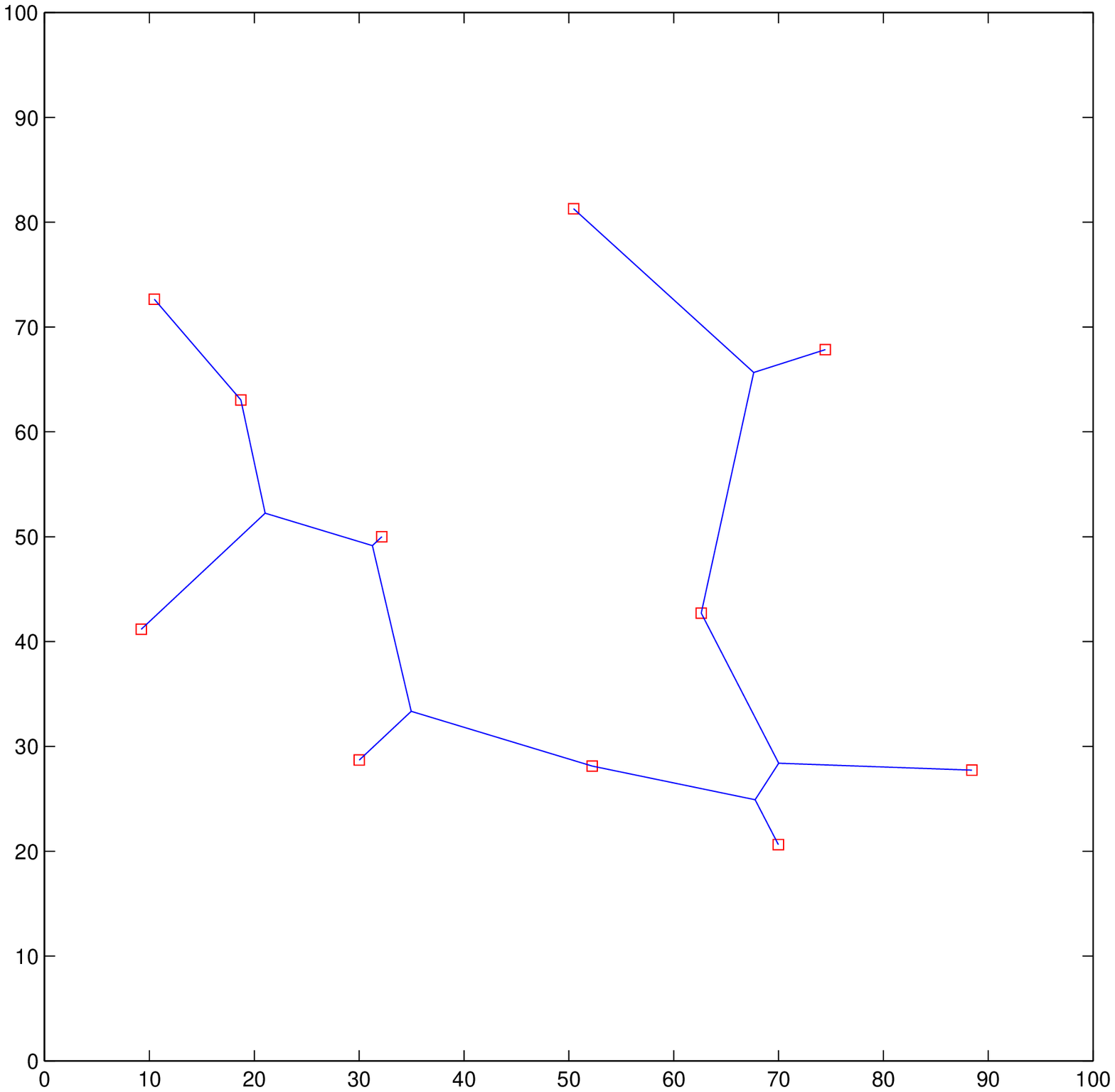}
\epsfxsize 7cm  
\epsfbox{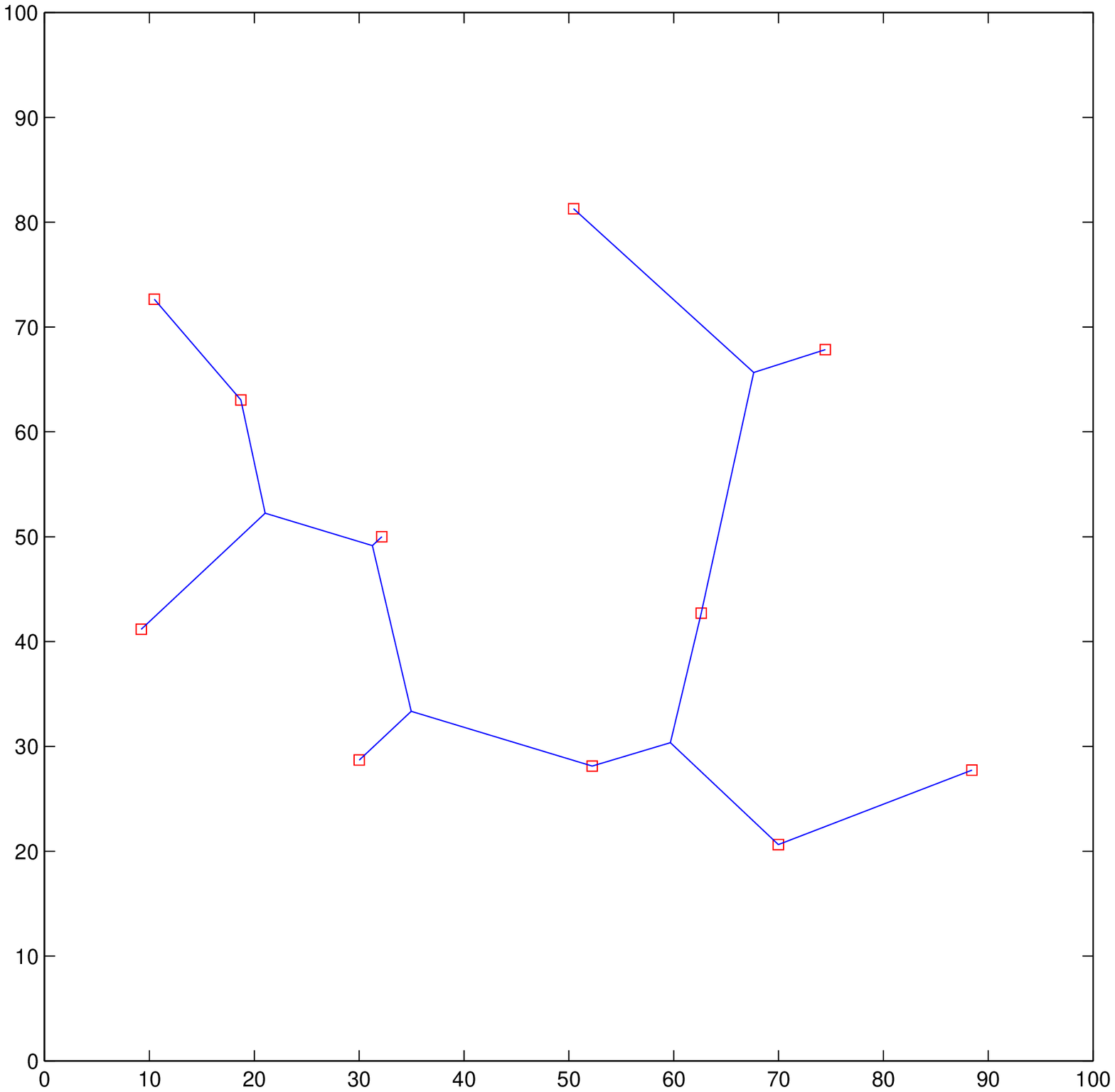}
\centerline{Figure 9: The completed Steiner tree.\hfill Figure 10: The supposedly minimal Steiner tree.}
\end{figure}

However, we performed several tests which suggest that the tree in Figure 10 is the shortest one. It has {\it Ltree} $=201.1988$.

\section{Full tree stretches}
\label{fts}

In the previous section we mentioned Figure 10, which contains a group of terminals connected by a full Steiner subtree. This subtree is obtained by the second option of the first mouse menu. Now we present some details and hints, so that the intuitive mind will hardly be wrong at identifying such groups.
\\

When will a subgroup of terminals admit connection by a full Steiner tree? For the time being, we answer this question providing each Steiner point is directly connected to a terminal of the subgroup (see Theorem 4.3 below). In general, it is when you have a ``good'' zigzag polygonal connecting them. For instance, run ``Mksaw'' for the terminal data {\tt test1.txt}, and do the same to ``stree''. You will get Figures 11 and 12.
\\
\input epsf
\begin{figure}[ht]
\epsfxsize 7cm  
\epsfbox{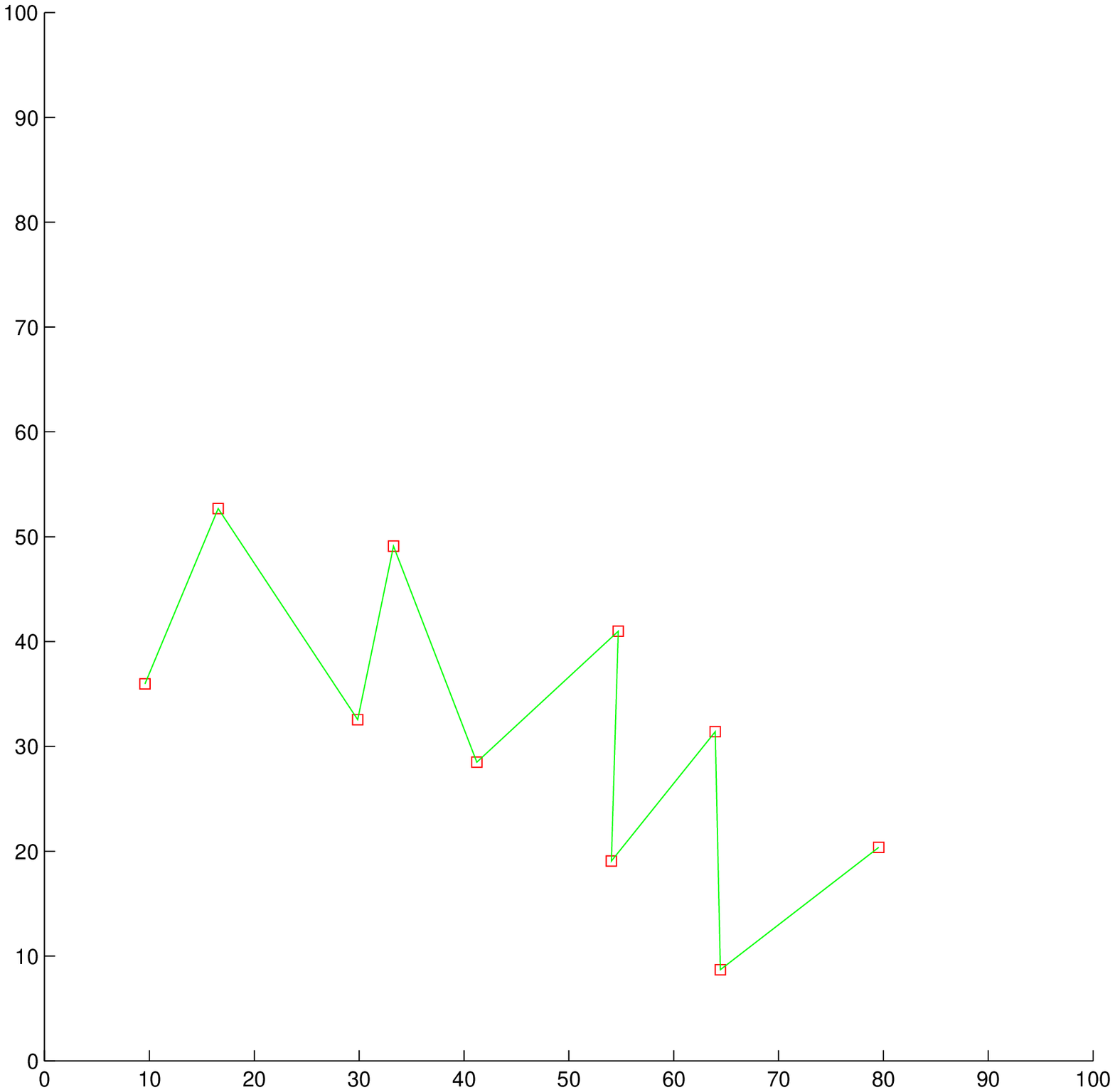}
\epsfxsize 7cm  
\epsfbox{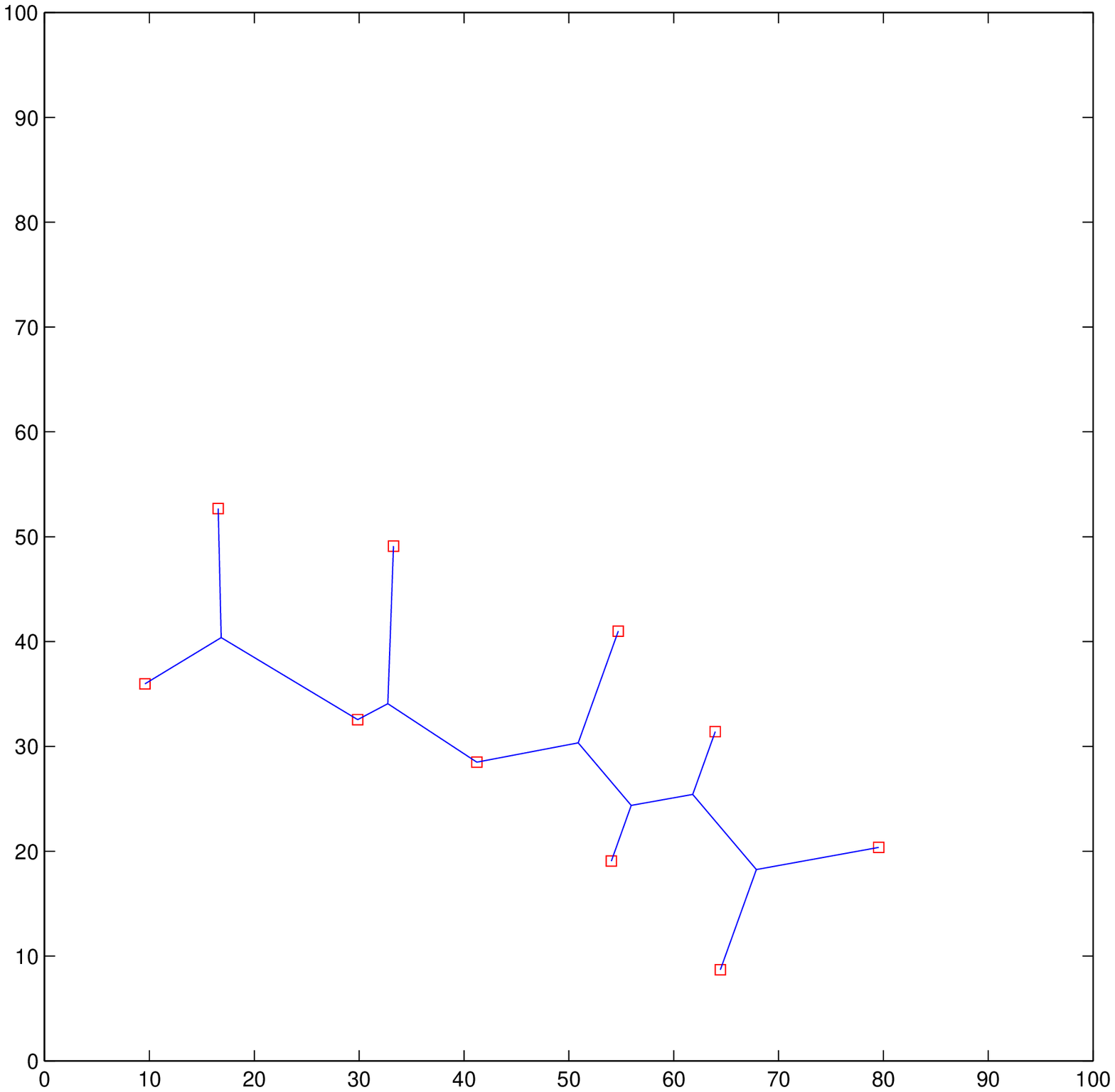}
\centerline{Figure 11: A ``good'' zigzag.\hfill Figure 12: An ``almost'' full tree.}
\end{figure}

This example shows that {\tt stree.m} will give an answer, even if no full tree exists. However, it relies on your guesses. Whenever they are wrong, choose the ``undo'' option from the second mouse menu, or simply run again ``stree'' if most of your terminals were a group like in Figure 11. This second case is illustrated in Figure 13 (run ``stree'' for {\tt test0} and omit the two uppermost terminals to get it). Figure 14 shows what is wrong: the zigzag is quite irregular. 
\\
\input epsf
\begin{figure}[ht]
\epsfxsize 7cm  
\epsfbox{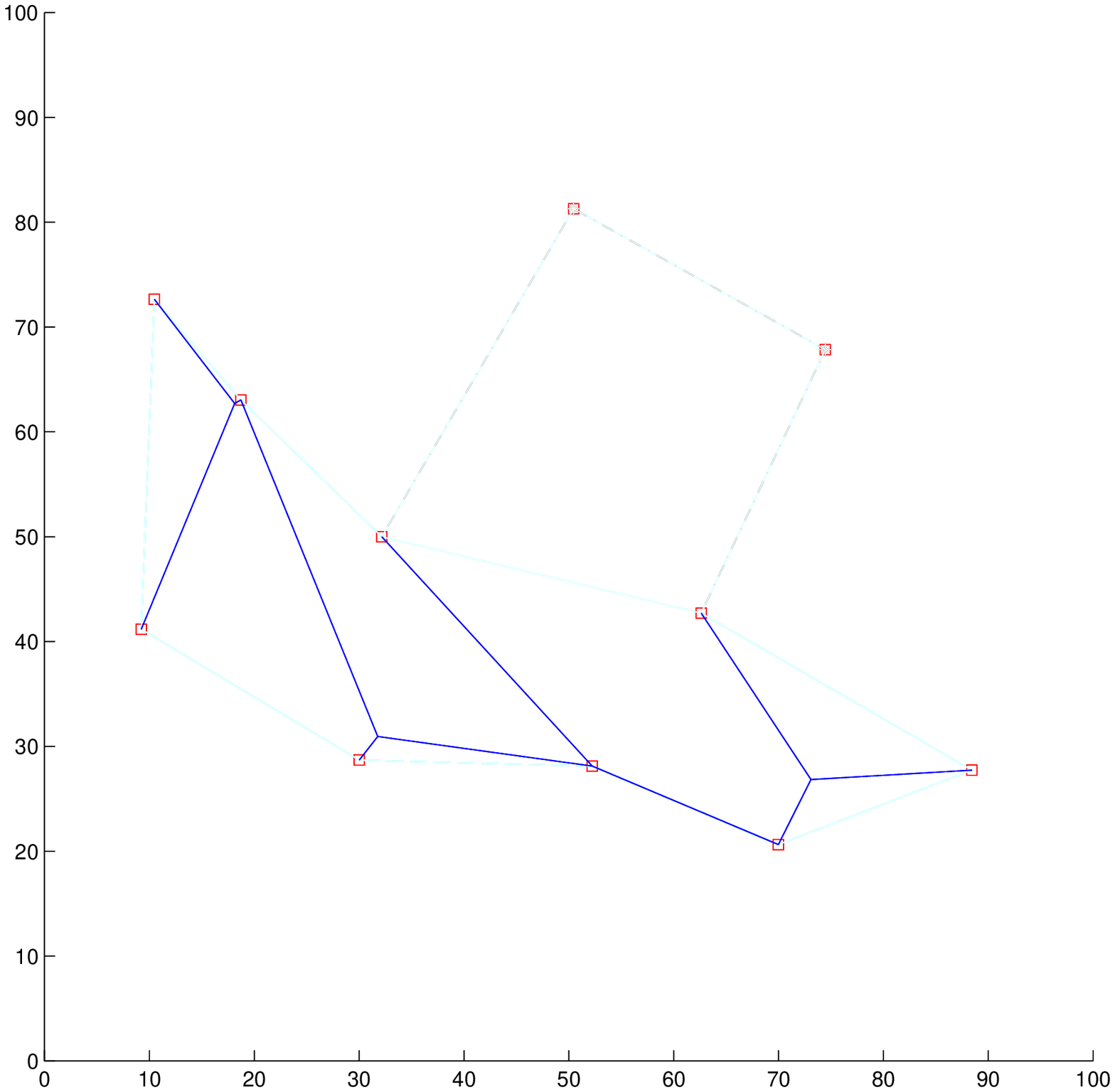}
\epsfxsize 7cm  
\epsfbox{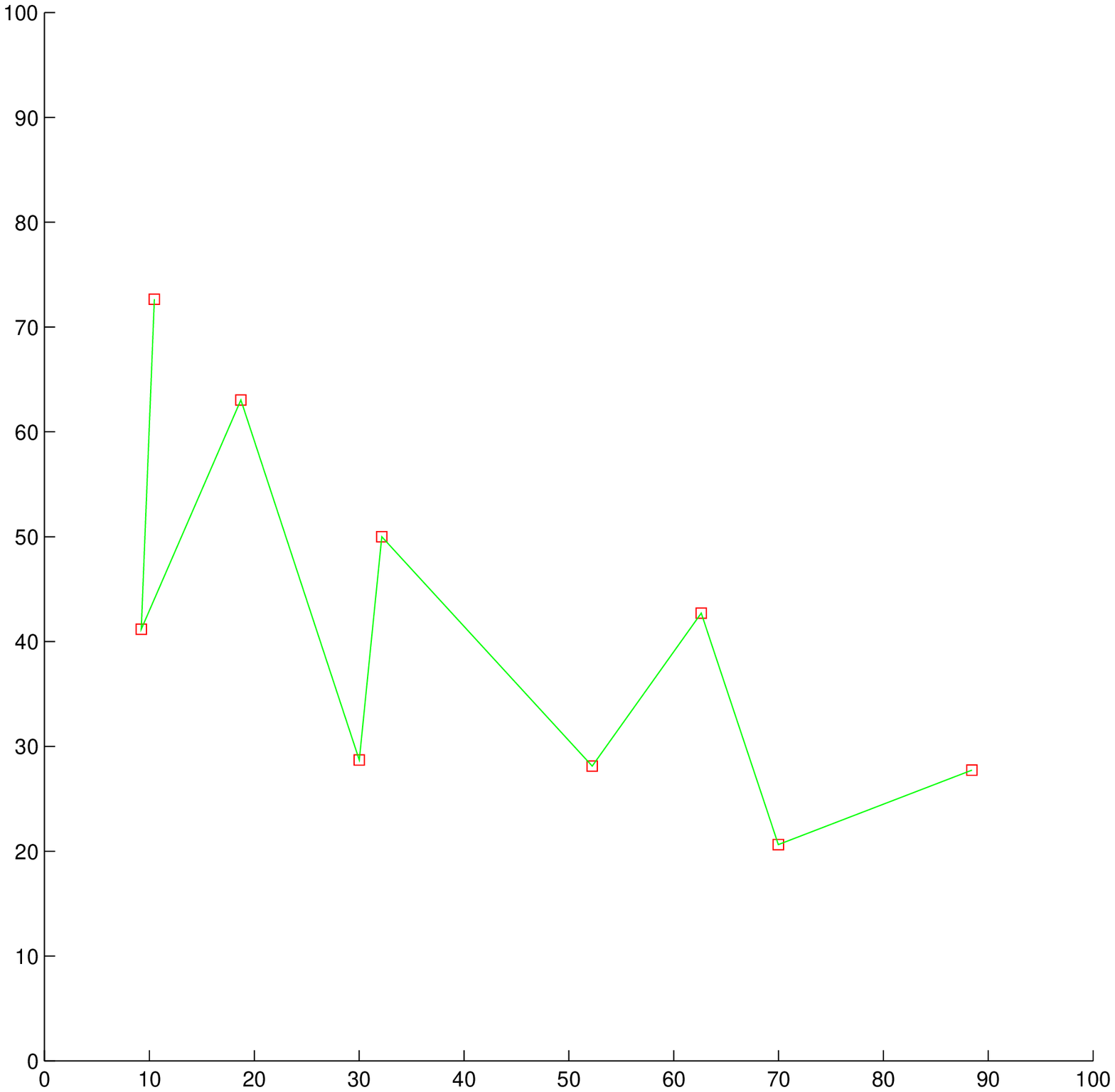}
\centerline{Figure 13: A wrong guess.\hfill Figure 14: A ``bad'' zigzag.}
\end{figure}

Here are some hints to identify good zigzags: the group of points should form a strip, which can scroll in any direction. The strip can be slightly bent or waved, but its width may oscillate even quite a lot. However, despite all hints we give, only the practice will really make you recognise the good zigzags very quickly. 
\\

The next section discusses some theorems, from Geometry and Complex Analysis, with which the programmes from {\tt stree.m} were written in order to obtain much shorter codes.

\section{Some theorems used in {\tt stree.m}}
\label{st}

The following theorems were used to implement {\tt Rprim.m} and {\tt Mksaw.m}, respectively. They all refer to the Euclidean distance.
\\
\\
{\bf Theorem 4.1.} {\it Let $S$ be a finite set of points with at least two elements. In this case, there are $P,\,Q\in S$ such that the distance between $P$ and $Q$ is maximal. The segment $PQ$ is called the {\it diameter} of $S$.}
\\
\\
{\bf Proof.} It is already explicitly given by lines 2-3 of the open code {\tt Rprim.m}, and so we shall omit it here.
\\
\\
{\bf Theorem 4.2.} {\it Let $Q$ be a quadrilateral with consecutive vertices $P_i$, $i=1,\cdots,4$. Let $vec_{i-2}=P_{i}-P_1$, $i\ge 2$, and $Vec_{i-1}=P_{i}-P_4$, $i\le 3$. Then $Q$ is convex precisely when $vec_1$ is between $vec_0$, $vec_2$, and also $Vec_1$ is between $Vec_0$, $Vec_2$.}
\\
\\
{\bf Proof.} By definition, $Q$ is convex exactly when $P_2$, $P_4$ are at opposite sides of $\reta{P_1P_3}$ and $P_1$, $P_3$ are at opposite sides of $\reta{P_2P_4}$. This is equivalent to the assertion of the Theorem.
\\
\\
{\bf Theorem 4.3.} {\it Let $A_1,\cdots,A_n$ be $n\ge3$ terminals in the complex plane $\C$ connected by a minimal full Steiner tree $S$. Let $V=\{V_1,\cdots,V_s\}$ be the set of Steiner points of $S$. Suppose that each element in $V$ admits a terminal such that both are the extremes of a segment in $S$. In this case, there exists $V_i\in V$ that determines all points in $V\setminus\{V_i\}$.} 
\\
\\
{\bf Proof.} By following the arguments from \S 3.4 of \cite{GP}, we have $s=n-2$ and only one segment in $S$ with extreme $A_k$, $\forall\,k\in\{1,\cdots,n\}$. Since $s\ge 1$, the arguments from \S 6 of \cite{GP} apply. Namely, if each element of $V$ were connected to a single terminal, then we would have $s=n$, a contradiction. Hence, there exists $V_i\in V$ that connects two terminals. Up to re-indexing, these are $A_1,A_2$ and $i=1$.

Now consider the ray given by 
\[
   r(t)=V_1+t\cdot\biggl(\frac{V_1-A_1}{|V_1-A_1|}+\frac{V_1-A_2}{|V_1-A_2|}\biggl),\,\,t\in\R_+.
\]

If $n=3$, then it is clear that a unique positive $\tau$ gives $r(\tau)=A_3$. Now take $n>3$. For each positive $t$, consider the rays $\rho_t=r(t)+(V_1-A_1)\cdot\R_+$ and $\varrho_t=r(t)+(V_1-A_2)\cdot\R_+$. Let $A=\{A_3,\cdots,A_n\}$ and take all positive $t$ such that $(\rho_t\cup\varrho_t)\cap A\ne\emptyset$. They will make a finite set $\{t_3,\cdots,t_m\}$ with $m\le n-2$. 

Since $V_1$ must be connected with another Steiner point, say $V_2$ (after re-indexing), then $V_2=r(t_k)$ for a certain $k\in\{1,\cdots,m\}$. Of course, $r(t_k)$ is connected to a terminal in $A$. Up to re-indexing, it is $A_k$.

In order to find $k$, we must repeat the same arguments with $V_1,A_k,r(t_k)$ respectively in the place of $A_1,A_2,V_1$, and so on. This will give at most $s$! full trees. One of them is minimal, whence all points $V_2,\cdots,V_s$ are determined. This concludes the proof of the Theorem.
\\

Our next section explains a bit of {\tt stree.m} and the open codes {\tt Rprim.m} and {\tt Mksaw.m}, which exemplify the applications of the theorems discussed in this present section. They correspond to {\tt rprim.p} and {\tt mksaw.p}, which are two of the files that {\tt stree.zip} contains.

\section{Efficient codes from Geometry and Complex Analysis}
\label{ec}

Figure 15 describes our pseudocode, which works recursively while the connection matrix is not complete. 
\\

{\bf {\tt stree} Pseudocode}

{\bf Input:} Set of terminals

{\bf Output:} Steiner tree

{\bf 1.} Pts $\gets$ terminals, Connexion\_Matriz $\gets$ 0 

{\bf 2.} Tree\_of\_Prim $\gets$ Compute\_Tree\_of\_Prim(Pts)

{\bf 3.} Steiner\_hull $\gets$ Compute\_Steiner\_hull(Pts)

{\bf 4.} While Connexion\_Matriz implies Stree non-connected

\hspace{1cm} {\bf i.} Subset\_Pts $\gets$ User's Choice of a Subgroup(Pts)

\hspace{1cm} {\bf ii.} Subtree $\gets$ Compute\_Connection(Subset\_Pts)

\hspace{1cm} {\bf iii.} If Subtree not OK, redo Step i

\hspace{1cm} {\bf iv.} Else Connexion\_Matriz $\gets$ Connection(Subset\_Pts)

{\bf 5.} return Steiner tree (and its length)
\input epsf
\begin{figure}[ht]
\centerline{Figure 15: The pseudocode of {\tt stree.m} algorithm.}
\end{figure}

Initially, {\tt stree.m} calls {\tt lune.m} and {\tt cvxhull.m} to inscribe the terminal points into the Steiner hull, namely a polygon coloured cyan as shown in Figure 4. In general, there will be isolated terminals inside the polygon. The ones that build its vertices are marked by {\tt stree.m} with either {\bf 0} or {\bf 1}, which mean ``greater'' and ``lesser'' than $120^\circ$, respectively. The characters {\bf 0} and {\bf 1} are stored in a string {\tt s}. For instance, if {\tt s} contains a stretch {\bf 010}, this hints to three terminals joined by a Steiner point. 
\\

But {\tt stree.m} does not connect them automatically, for this hint might not lead to the shortest tree. We use the variable {\tt s} for purposes like building zigzags out of stretches {\bf 01...10}. Of course, not all zigzags are good, but the programme will try them back and forth, and even split them in parts. We are certain that, if you have something like Figure 13, then a supervised splitting will work better than many automatic tests.
\\

Of course, {\tt lune.m} and {\tt cvxhull.m} are based upon the Convex Hull and Lune properties described in \cite{GP}. They do not give a polygon with zigzags, but with stretches like in Figure 16. Hence, {\tt stree.m} calls {\tt mksaw.m} in order to get the zigzag. 
\\
\input epsf
\begin{figure}[ht]
\epsfxsize 7cm  
\epsfbox{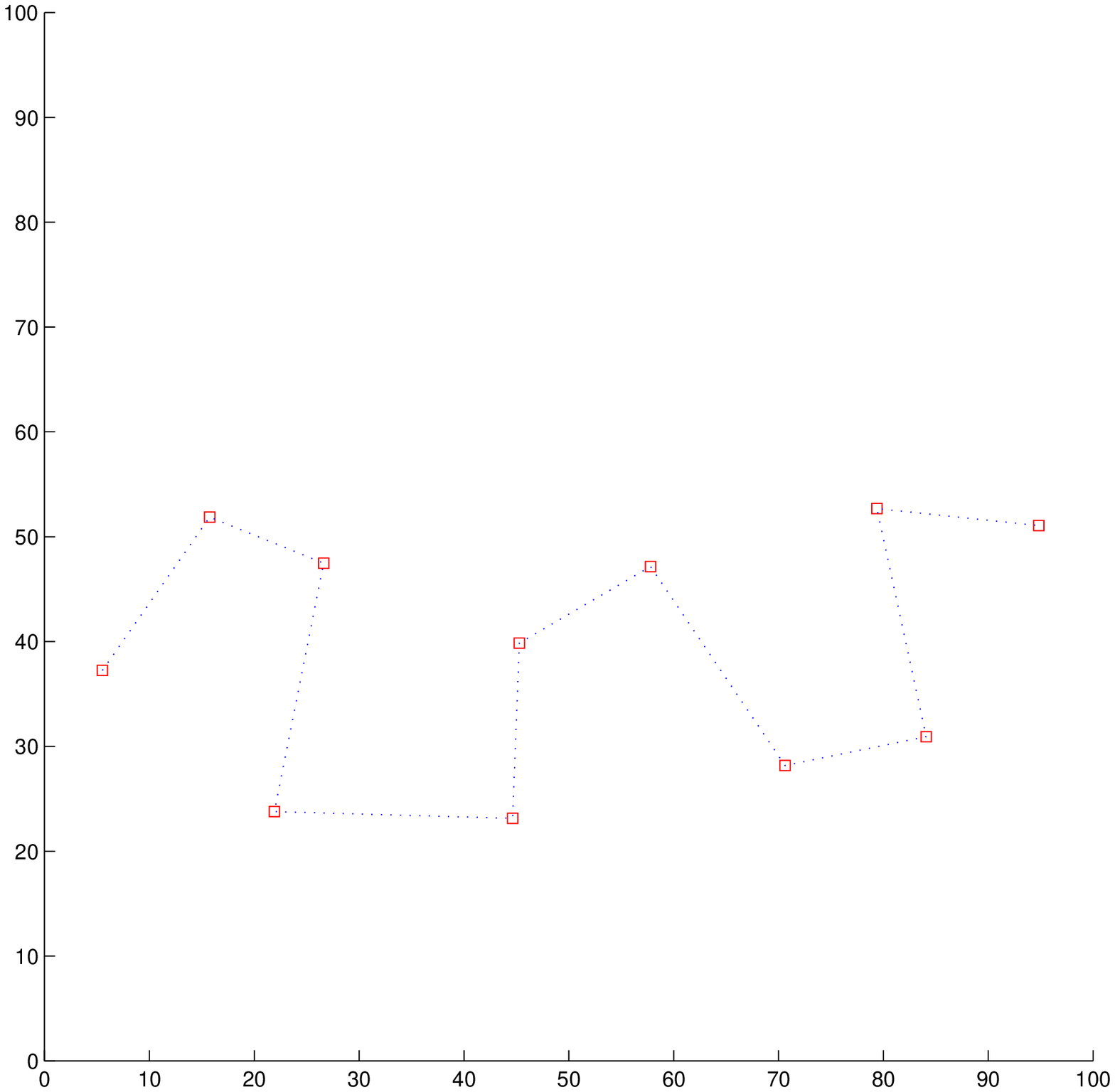}
\epsfxsize 7cm  
\epsfbox{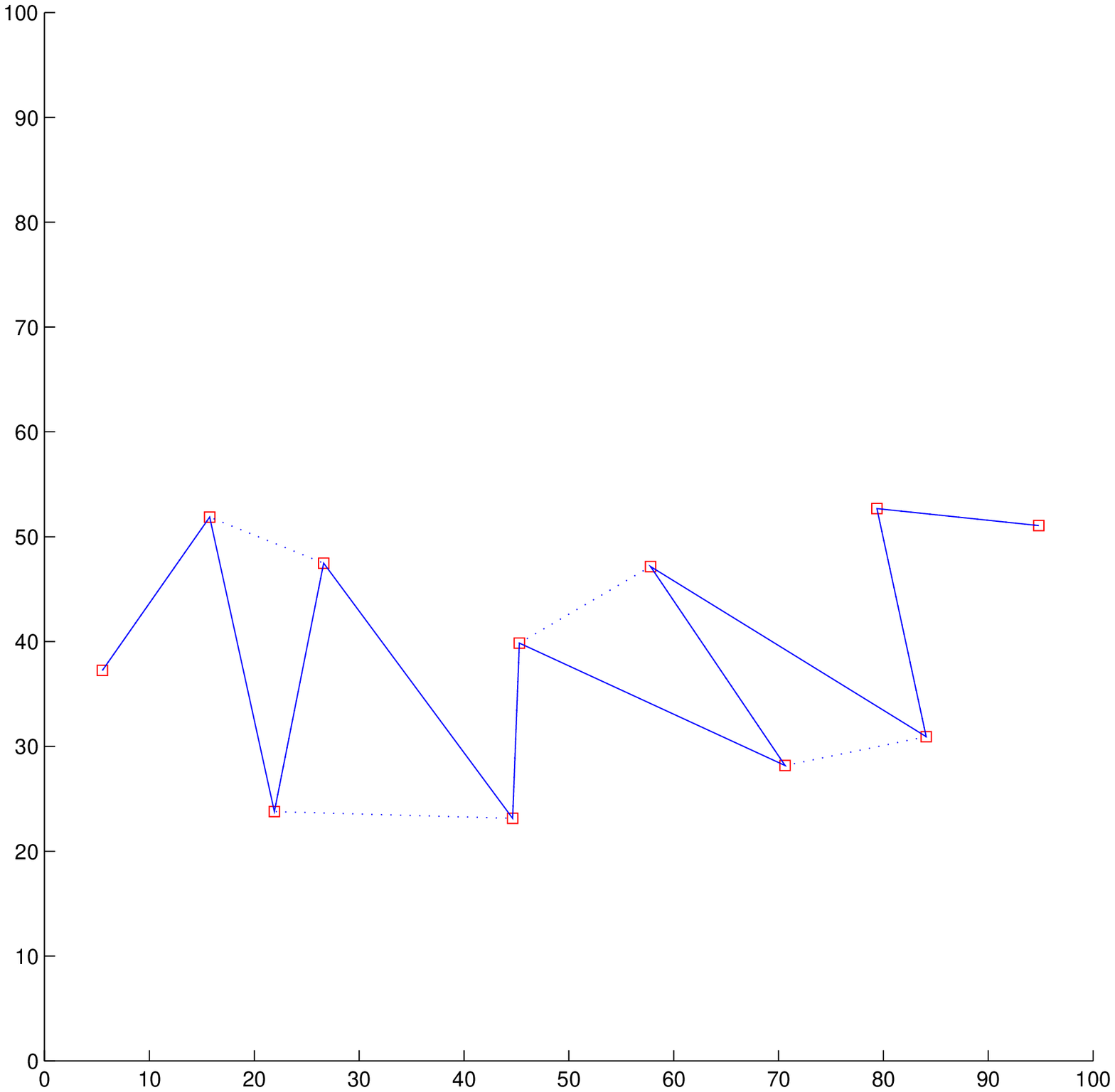}
\centerline{Figure 16: A ``{\bf 01\dots 10}'' stretch.\hfill Figure 17: Its rearrangement by ``Mksaw''.}
\end{figure}

If you want to test these procedures, run ``Cvxhull'' for the terminal data {\tt test2.txt} and save the output with the name {\tt test3}. A figure will show the original input order of the terminals (dotted lines), and a blue polygon involving them (the convex hull). Then run ``Lune'' for {\tt test3.txt} and name the output {\tt test4}. 
\\

Now we are going to explain the open codes {\tt Rprim.m} and {\tt Mksaw.m}, which will finally generate a zigzag out of {\tt test2.txt} (the initial datafile). According to Theorem 4.1, there is a diameter for the points in {\tt test2.txt}, and lines 2-3 of {\tt Rprim.m} are precisely the commands to find them, namely {\tt Pts(h(k))} and {\tt Pts(k)}. However, you {\it must} run ``Rprim.m'' for the re-ordered data {\tt test4}. 
\\

Now we project all points along the diameter and re-order them by increasing distance between their projection and the extreme {\tt Pts(k)}. That is what the while-loop does in lines 11 to 19. The re-ordered input is restored in line 20, and may be saved in line 21. We did it in {\tt test5.txt}, which {\tt Mksaw.m} finally re-orders into a zigzag.
\\

Notice that the terminals from {\tt test5.txt} are in zigzag order {\it except} for one single square wavelet. However, the open code {\tt Mksaw.m} works well even if you have a totally square wave. Enter the terminal points in an order that makes such a wave. We have used {\tt test6.txt} to generate Figure 16. 
\\

You have just seen that this precaution is not necessary when you run {\tt stree.m}, for it will rearrange the points as explained above. 
\\

By walking along the square wave from the very first point on, two consecutive steps will always lead to a vertex that makes a quadrilateral $Q$ with the following two. That is why the while-loop ends in {\tt pts=pts(2:end)}.
\\

Now look at lines 4 to 9 of {\tt Mksaw.m} and apply Theorem 4.2 to $Q$. The symbol $\times$ will indicate the {\it vector product}. From Geometry, the convexity criteria given by this theorem is equivalent to $vec_0\times vec_1$, $vec_2\times vec_1$ pointing in opposite directions. The same holds for $Vec_0\times Vec_1$, $Vec_2\times Vec_1$. Taking these vectors as complex numbers in $\C\times\R$, say $vec_1=a+ib=(a,b,0)$ and $vec_0=c+id=(c,d,0)$, we have $vec_0\times vec_1=(0,0,bc-ad)$, namely $bc-ad=Im\{vec_1\cdot\ovl{vec}_0\}$. Thus, from Complex Analysis the vectors $vec_0\times vec_1$, $vec_2\times vec_1$ will point in opposite directions precisely when the signs of $Im\{vec_1\cdot\ovl{vec}_0\}$ and $Im\{vec_1\cdot\ovl{vec}_2\}$ are opposite. Hence, lines 10-11 from {\tt Mksaw.m} check if $Q$ is convex. In this case, the programme makes a sawtooth out of $Q$. Then we go two steps forward with the command {\tt pts=pts(2:end);} and the while-loop repeats the process, unless we have already came to the end of the line.

\section{Conclusions}
\label{ccls}

Differently from the approach of trying a fully automated method, we propose to take advantage of the good choices that a user can make. Many attributes, like intuition, guessing, practice and sometimes a bird's-eye view, are valuable means that one cannot translate into any programming language. It is just because these attributes are chiefly {\it human}. Hence, as long as a task is feasible with the help of supervision, we suggest taking it into account, besides the fully automated methods. This proposal is not new, but we endeavour to obtain a code that is both easy to run and to understand.
\\

The programme {\tt stree.m} is still in the beta version. Further improvements will include more feedback to the user. For instance, the tree will be checked with the (Double) Wedge and Diamond properties, among others (see \cite{GP} for details). Some tests can be implemented to run {\it while} the users draw, so that they may also undo steps which just {\it seem} successful with this present version.   
\\

Moreover, {\tt stree.m} still works strongly devoted to {\it real} Steiner trees, which in fact should be adapted to practical purposes. For instance, outputs consider even Steiner points extremely close to a terminal. By implementing such a tree to a multicast network, those Steiner points are unnecessary, and even costly. In future, the user will decide on the tolerance regarding the minimum distance that terminals and Steiner points will keep apart. 
\\

By the way, it is even preferable to implement multicast networks with a minimum number of Steiner points, because of the high cost of the routers. This is also the case of WDM optical networks (see \cite{CD}). Therefore, it will be useful to have future versions of {\tt stree.m} devoted to the construction of such trees. The {\it rectilinear} Steiner trees are also of interest (see \cite{FK,Z}), and then another option like {\tt stree.m} to be developed.
\\

\noindent \Large\textbf{Acknowledgment}\\[2mm] 

\footnotesize{Many improvements in this paper were due to the careful analyses carried out by referees. We thank them for their valuable help.}\\[3mm]

\end{document}